\documentclass[sigconf, nonacm]{acmart}

\renewcommand\footnotetextcopyrightpermission[1]{}
\usepackage{multirow}
\usepackage{enumitem}

\AtBeginDocument{%
  }

\definecolor{amber}{rgb}{1.0, 0.49, 0.0}
\definecolor{dodgerblue}{RGB}{30, 144, 255}
\definecolor{violet}{RGB}{238,130,238}
\definecolor{my_green}{RGB}{113,173,71}
\definecolor{my_blue}{RGB}{44,115,182}

\newcommand{\reffig}[1]{\textcolor{black}{Figure~\ref{fig:#1}}} 
\newcommand{\refsec}[1]{\textcolor{black}{Section~\ref{sec:#1}}}
\newcommand{\reftab}[1]{\textcolor{black}{Table~\ref{tab:#1}}}

\newcommand{\eg}[1]{\textcolor{black}{\textit{e.g.,~}}}
\newcommand{\ie}[1]{\textcolor{black}{\textit{i.e.,~}}}

\begin{document}

\title{Style Customization of Text-to-Vector Generation with Image Diffusion Priors}


\author{Peiying Zhang}
\affiliation{
 \institution{City University of Hong Kong}
  \city{Hong Kong}
  \country{China}
 }
\email{zhangpeiying17@gmail.com}

\author{Nanxuan Zhao}
\affiliation{
 \institution{Adobe Research}
  \city{San Jose}
  \country{USA}
 }
\email{nanxuanzhao@gmail.com}

\author{Jing Liao}
\authornote{Corresponding author}
\affiliation{
 \institution{City University of Hong Kong}
 \city{Hong Kong}
 \country{China}
 }
\email{jingliao@cityu.edu.hk}

\begin{abstract}

Scalable Vector Graphics (SVGs) are highly favored by designers due to their resolution independence and well-organized layer structure. Although existing text-to-vector (T2V) generation methods can create SVGs from text prompts, they often overlook an important need in practical applications: style customization, which is vital for producing a collection of vector graphics with consistent visual appearance and coherent aesthetics.

Extending existing T2V methods for style customization poses certain challenges. Optimization-based T2V models can utilize the priors of text-to-image (T2I) models for customization, but struggle with maintaining structural regularity. On the other hand, feed-forward T2V models can ensure structural regularity, yet they encounter difficulties in disentangling content and style due to limited SVG training data. 

To address these challenges, we propose a novel two-stage style customization pipeline for SVG generation, making use of the advantages of both feed-forward T2V models and T2I image priors. In the first stage, we train a T2V diffusion model with a path-level representation to ensure the structural regularity of SVGs while preserving diverse expressive capabilities. In the second stage, we customize the T2V diffusion model to different styles by distilling customized T2I models. By integrating these techniques, our pipeline can generate high-quality and diverse SVGs in custom styles based on text prompts in an efficient feed-forward manner. The effectiveness of our method has been validated through extensive experiments. The project page is https://customsvg.github.io.

\end{abstract}


\begin{teaserfigure}
  \includegraphics[width=\textwidth]{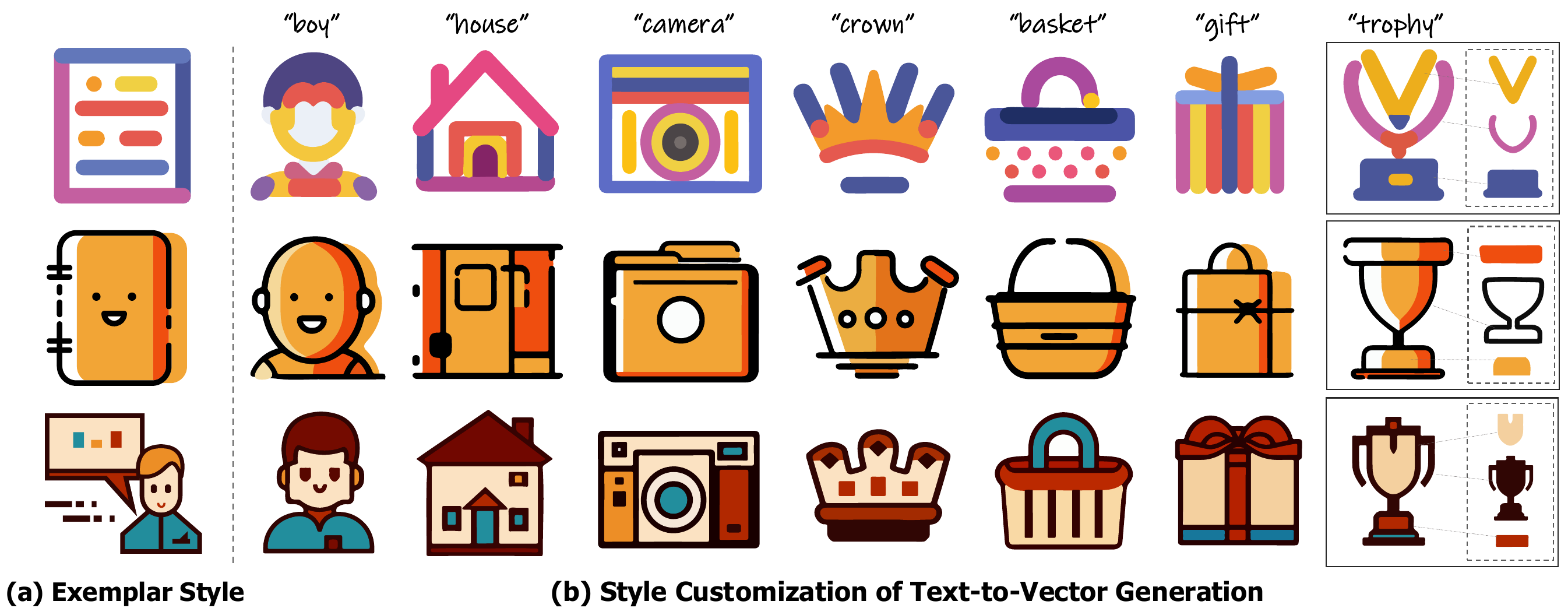}
  \caption{ Examples of vector graphics generated from text prompts in custom styles using our method, showcasing structural regularity and expressive diversity. Exemplar SVGs: the $1^{st}$ and $3^{rd}$ rows are from \copyright{SVGRepo}; the $2^{nd}$ row is from \copyright{iconfont}. }
  \label{fig:teaser}
\end{teaserfigure}

\maketitle

\section{Introduction}
\label{sec:introduction}

Vector graphics, especially in the form of Scalable Vector Graphics (SVG), play an essential role in digital arts such as icons, clipart, and graphic design. By representing visual elements as geometric shapes, SVGs provide resolution independence, compact file sizes, and flexibility for layer-wise manipulation, making them highly favored by designers. 
Given the challenges of creating high-quality vector graphics, many recent works \cite{wu2023iconshop, jain2022vectorfusion, thamizharasan2024vecfusion, xing2024svgfusion} have proposed algorithms in text-to-vector (T2V) generation. However, these methods overlook an important need in practical applications - style customization. Designers often customize a set of vector graphics with consistent visual appearance and aesthetic coherence. This is crucial for ensuring design quality, particularly in contexts like branding, user interfaces, and themed illustrations.

Simply extending existing T2V methods for style customization is difficult. Current T2V methods can be categorized into optimization-based and feed-forward methods.
Optimization-based T2V methods, which either optimize a set of vector elements (\eg, cubic Bézier curves) to fit the images generated by T2I models \cite{zhang2023text, ma2022towards}, or directly optimize shape parameters using Score Distillation Sampling (SDS) loss \cite{poole2022dreamfusion} based on T2I models \cite{zhang2024text, jain2022vectorfusion, xing2023svgdreamer}, can be extended for style customization by fine-tuning a T2I model on user-provided style examples. Although effective in adapting to new styles, these methods are time-consuming and often produce fragmented or cluttered paths. Such outputs overlook the inherent structural regularity and element relationships within SVG designs, making them difficult to edit or refine further.

Feed-forward T2V methods, on the other hand, are trained on SVG datasets using large language models (LLMs) \cite{wu2023iconshop, rodriguez2023starvector} or diffusion models \cite{thamizharasan2024vecfusion, xing2024svgfusion}, maintaining SVG regularity and attaining high-quality outcomes within their respective training domains.
However, style customization of the T2V model presents significant challenges. The absence of large-scale, general-purpose text-SVG datasets makes it difficult for the T2V model to disentangle content and style semantics, limiting its ability to generalize to new styles. Consequently, a straightforward approach of fine-tuning a base T2V model with only a few exemplar SVGs, following T2I customization techniques \cite{ruiz2022dreambooth, hu2021lora, kumari2022multi}, often leads to overfitting on the exemplar SVGs. Nevertheless, acquiring a sufficient number of consistent style sample SVGs for fine-tuning is impractical due to the scarcity of such data.

Addressing these limitations, we propose a novel two-stage style customization pipeline for SVG generation using only a few exemplar SVGs. It combines the strengths of feed-forward T2V methods to ensure SVG structural regularity and T2I models to acquire powerful customization capabilities. In the first stage, we train a T2V model on black-and-white SVG datasets to focus on learning the contents and structures of SVGs. In the second stage, we learn various styles of SVGs by distilling priors from different customized T2I models. Our two-stage pipeline also helps the T2V model explicitly disentangle content and style semantics.

The aim of the first stage is to train a T2V generative model tailored for style customization. Considering that LLM-based methods generate SVG code in an autoregressive manner, limiting their ability to utilize raster images as supervision, we adopt a diffusion model as the base model to enable customization from the T2I model. As for the representation, global SVG-level representations \cite{xing2024svgfusion} suffer from limited expressivity constrained by the dataset \cite{rombach2022high}, and point-level representations \cite{thamizharasan2024vecfusion} are inefficient for complex SVGs. Thus, we select a path-level representation \cite{zhang2024text} that ensures both compactness and expressivity. In this stage, our path-level T2V diffusion model learns to generate SVGs that feature text-aligned content and exhibit structural regularity.

In the second stage, we distill styles from a T2I diffusion model to enable style customization for the T2V diffusion model. Specifically, we fine-tune the T2I model using a small set of style images to generate diverse customized images, which serve as augmented data for training the T2V model. To facilitate image-based training, we employ a reparameterization technique \cite{song2020score} to compute SVG predictions and render them as images, enabling the T2V model to be updated via an image-level loss. After training, the T2V model can generate SVGs in learned custom styles corresponding to text prompts in a feed-forward manner.

We evaluate our method through comprehensive experiments across vector-level, image-level, and text-level metrics. The results demonstrate the effectiveness of our model in generating high-quality vector graphics with valid SVG structures and diverse customized styles, given input text prompts. Examples of style customization results produced by our framework are shown in \reffig{teaser}. Our key contributions are:
\begin{itemize}[leftmargin=*]
  \item We propose a novel two-stage T2V pipeline to disentangle content and style in SVG generation, which is also the first feed-forward T2V model capable of generating SVGs in custom styles.
  \item  We design a T2V diffusion model based on path-level representations, ensuring structural regularity of SVGs while maintaining diverse expressive capabilities.
  \item We develop a style customization method for the T2V model by distilling styles from customized image diffusion models.
\end{itemize}

\begin{figure*}[tbp]
  \centering
  \includegraphics[width=1.0\linewidth]{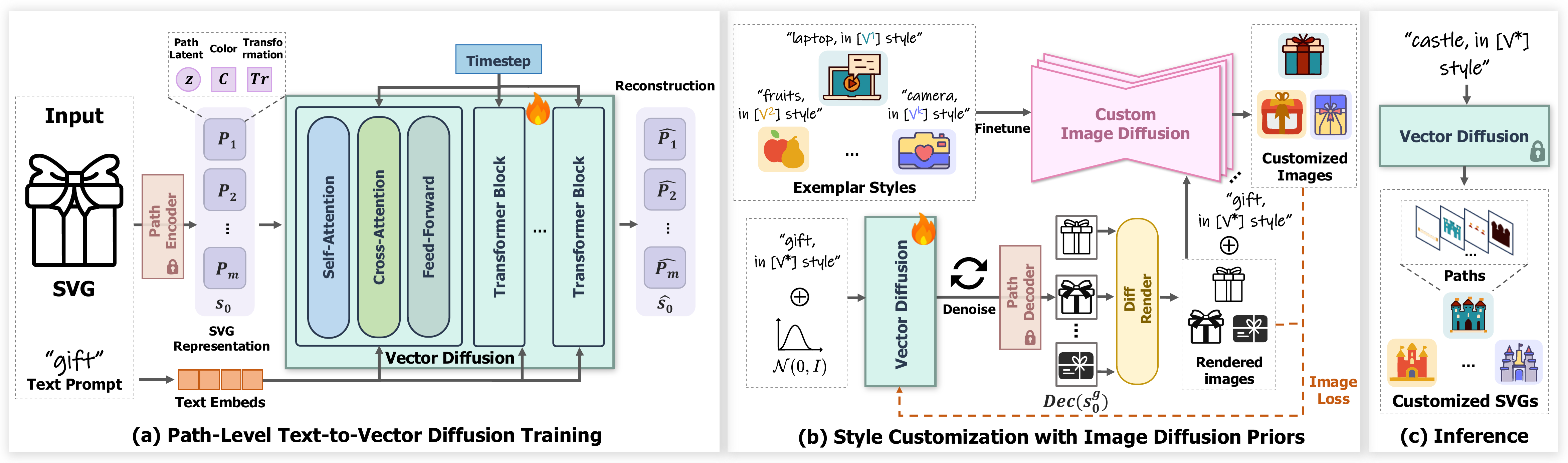}
  \caption{ \label{fig:pipeline} Our two-stage style customization pipeline for SVGs. (a) In Stage 1, we train a path-level T2V diffusion model on black-and-white SVG datasets to focus on learning the contents and structures of SVGs. (b) In Stage 2, we learn various styles of SVGs by distilling priors from different customized T2I models. (c) After training, our T2V model can generate SVGs in custom styles learned during Stage 2 in a feed-forward manner by appending the corresponding style token to the text prompt. Exemplar SVGs are from \copyright{SVGRepo}. }
\end{figure*}

\section{Related Work}
\label{sec:related_work}

\subsection{Optimization-based T2V Generation}
\label{sec:optimization_based_t2v_generation}

Optimization-based methods leverage pre-trained vision-language models, such as CLIP \cite{radford2021learning} or diffusion models \cite{rombach2022high}, combined with differentiable rendering \cite{li2020differentiable} to directly optimize SVG paths.
CLIP-based methods \cite{frans2022clipdraw, schaldenbrand2022styleclipdraw, song2022clipvg, vinker2022clipasso} maximize image-text alignment within CLIP latent space.
Recent works exploit Score Distillation Sampling (SDS) loss \cite{poole2022dreamfusion, wang2023prolificdreamer} to capitalize on the strong visual and semantic priors of T2I diffusion models. These methods can produce static \cite{jain2022vectorfusion,iluz2023word,xing2023diffsketcher,xing2023svgdreamer,zhang2024text} or animated \cite{gal2023breathing, wu2024aniclipart} SVGs aligned with text prompts. However, each optimization typically requires tens of minutes per SVG, making them impractical for real design scenarios.
Alternatively, some commercial tools \cite{adobe-illustrator, illustroke} integrate T2I models with vectorization techniques \cite{kopf2011depixelizing, selinger2003potrace, favreau2017photo2clipart, hoshyari2018perception, dominici2020polyfit, ma2022towards} to convert raster images into SVGs.
Despite their visually appealing results, these methods often include multiple fragmented paths and lack coherent layer relationships in SVGs, complicating further edits.
While some methods \cite{zhang2023text, warner2023interactive} adapt paths from an exemplar SVG via semantic correspondences to preserve layer structure, they are unsuitable for style customization when the source and target differ significantly in semantics.

\subsection{Feed-forward T2V Generation}
\label{sec:feed_forward_t2v_generation}

Feed-forward methods have explored to learn SVG properties from specialized datasets using large language models or diffusion models.
LLM-based approaches \cite{wu2023iconshop, tang2024strokenuwa, rodriguez2023starvector} treat SVG scripts as text by designing special tokenization schemes, allowing command sequences to be combined with text tokens in an autoregressive manner.
Diffusion-based methods \cite{wang2023sketchknitter, thamizharasan2024vecfusion, xing2024svgfusion} design various SVG representations and model architectures within the vector domain.
Although these feed-forward pipelines are conceptually elegant, their generalization capabilities are constrained by the absence of large-scale, general-purpose vector graphics datasets.
Consequently, they are limited to producing SVGs in fixed styles, while our approach supports diverse customized styles in a feed-forward manner.

\subsection{Cusomization of T2I Generation}
\label{sec:cusomization_of_t2i_generation}

Recent advances in T2I customization have enabled flexible adaptation of concepts and styles using only a few reference images \cite{gal2022image, ruiz2022dreambooth, kumari2022multi}. The pioneering approach DreamBooth \cite{ruiz2022dreambooth} fine-tunes the entire diffusion model by associating user-provided concepts with a unique token. Parameter-efficient fine-tuning (PEFT) methods \cite{peft} propose modifying only specific network components, such as low-rank weight offsets \cite{hu2021lora, frenkel2025implicit}, cross-attention blocks \cite{kumari2022multi, ye2023ip}, or adapter layers \cite{sohn2023styledrop, mou2024t2i}. 
While effective for customizing powerful T2I diffusion models, these techniques are challenging to apply to T2V models, which have limited generalization ability and struggle to disentangle content and style semantics, often leading to overfitting when fine-tuned with only a few exemplar SVGs.
Our two-stage style customization pipeline addresses these limitations by leveraging T2I diffusion priors to help the T2V model learn various styles of SVGs.

\section{Overview}
\label{sec:overview}

Our goal is to generate SVGs that are customized to specific styles while aligning with the semantics of given text prompts and maintaining structural regularity. To achieve this, we propose a novel two-stage style customization pipeline designed to disentangle content and style in SVG generation. An illustration of the pipeline is shown in \reffig{pipeline}.

\paragraph{Path-Level Text-to-Vector Diffusion Training (\refsec{text_to_vector_diffusion_model})}
In the first stage, we train a T2V model that focuses on learning the contents and structures of SVGs. We adopt a path-level representation that ensures both compactness and expressivity, and train this path-level T2V diffusion model on black-and-white SVG datasets. 

\paragraph{Style Customization with Image Diffusion Priors (\refsec{style_customization_with_image_diffusion_priors})}
In the second stage, we aim to customize the T2V model to generate SVGs in diverse new styles with only a few exemplars. We fine-tune the T2I diffusion model on a small set of style images to produce diverse customized images, which are used as augmented data to train the T2V model through an image-level loss.

\section{Path-Level Text-to-Vector Diffusion Training} 
\label{sec:text_to_vector_diffusion_model}

In the first stage, we train a T2V diffusion model to generate SVGs aligned with text semantics while ensuring structural regularity. To achieve this, we adopt a compact and expressive path-level representation, and train the model on black-and-white datasets, focusing on learning SVG content and structure.

\subsection{SVG Representation}
\label{sec:svg_representation}

An SVG can be represented as a set of paths, denoted as $SVG = \{Path_1, Path_2, \ldots, Path_m\}$. A parametric path can be defined as a series of cubic Bézier curves connected end-to-end and filled with a uniform color $c$, represented as $Path_i = (p_1, p_2, \ldots, p_d, c)$, where $\{p_j\}_{j=1}^d$ are the $d$ control points used to define the cubic Bézier curves.
In contrast to recent approaches that use global SVG-level representations \cite{xing2024svgfusion}, which are constrained in expressivity by the limitations of the SVG dataset, or point-level representations \cite{thamizharasan2024vecfusion}, which become inefficient for complex SVGs, we adopt a path-level representation that balances compactness and expressivity.

T2V-NPR \cite{zhang2024text} introduced a path-level SVG VAE designed to effectively capture common shape patterns and geometric constraints within its latent space, ensuring smooth path outputs.
Following the methodology of T2V-NPR, we leverage a pre-trained SVG VAE to encode the $d$ control points of each path into a latent vector $z_i$. This latent vector is then combined with the associated color $C_i$ and transformation parameters $Tr_i$ for the $i$-th path, denoted as $\mathbf{P}_i = (z_i, C_i, Tr_i)$. Using this path-level representation, an SVG tensor can be represented as a sequence of $m$ paths in the latent space, denoted as $\mathbf{s}_0 = (\mathbf{P}_1, \mathbf{P}_2, \ldots, \mathbf{P}_m)$, where $\mathbf{s}_0 \in \mathbb{R}^{d_P \times m}$ and $d_P$ is the dimension of the path embeddings.

\subsection{Vector Denoiser}

The vector denoiser is trained to reverse a Gaussian diffusion process, enabling the generation of SVG tensors from noisy inputs. In the forward diffusion process, Gaussian noise is progressively added to a sample SVG tensor $\mathbf{s}_0$ over $T$ time steps, ultimately transforming it into a unit Gaussian noise $ \mathbf{s}_T \sim \mathcal{N}(0, \mathbf{I})$ \cite{ho2020denoising}. At each time step $t \in \{1, 2, \ldots, T\}$, the vector denoiser predicts the noise content $\boldsymbol{\epsilon}_\theta$ to be removed from the noisy SVG representation $\mathbf{s}_t$.

\paragraph{Model Architecture}
We adopt a transformer architecture based on DiT \cite{peebles2023scalable} as the backbone for our vector denoiser. As shown in \reffig{pipeline}(a), the model takes the noisy tensor $\mathbf{s}_t$ as input and is conditioned on both the text prompt $\mathbf{y}$ and the time step $t$. Each transformer block consists of self-attention, cross-attention, and feed-forward layers.
The text prompt $\mathbf{y}$ is encoded into feature embeddings using the CLIP text encoder \cite{radford2021learning}, which interact with vector features through cross-attention similar to \cite{chen2023pixart}. Time step embeddings are injected via adaptive layer normalization \cite{chen2024towards} in each transformer block.

\paragraph{Training Objective}

The training of the T2V diffusion model follows the denoising diffusion probabilistic model (DDPM) framework \cite{ho2020denoising}. At each time step $t$, Gaussian noise $\boldsymbol{\epsilon} \sim \mathcal{N}(0, \mathbf{I})$ is added to the original SVG tensor $\mathbf{s}_0$, resulting in a noisy representation $\mathbf{s}_t = \sqrt{\bar{\alpha}_t} \mathbf{s}_0 + \sqrt{1 - \bar{\alpha}_t} \boldsymbol{\epsilon}$, where $\bar{\alpha}_t$ is a time-dependent coefficient controlling the noise level.
The vector denoiser is trained to predict the added noise $\boldsymbol{\epsilon}$ in $\mathbf{s}_t$, conditioned on the text prompt $\mathbf{y}$ and the time step $t$. The training objective is to minimize the $L_2$ distance between the predicted noise $\boldsymbol{\epsilon}_\theta$ and the actual noise $\boldsymbol{\epsilon}$:
\begin{equation}
    \mathcal{L}_{\text{DM}} = \mathbb{E}_{\mathbf{s}_0, \mathbf{y}, \boldsymbol{\epsilon}, t} \left[ \left\| \boldsymbol{\epsilon} - \boldsymbol{\epsilon}_\theta(\mathbf{s}_t, t, \mathbf{y}) \right\|_2^2 \right].
    \label{eq:diffusion_loss}
\end{equation}

\subsection{Training Details}

\begin{figure}[tbp]
  \centering
  \includegraphics[width=1.0\columnwidth]{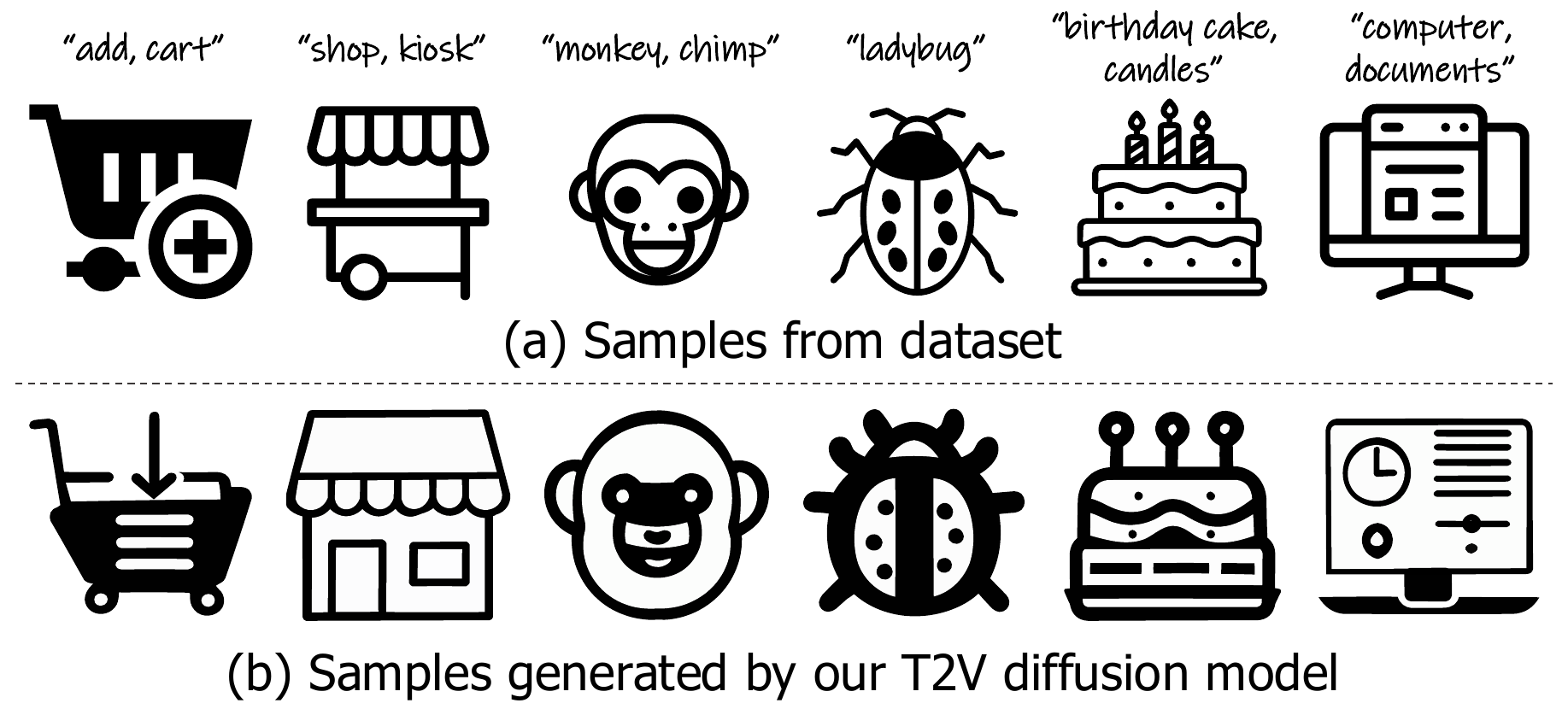}
  \caption{ \label{fig:vector_diff_samp} (a) SVG examples from the dataset. (b) SVG samples generated from random noise by our T2V diffusion model in Stage 1. }
\end{figure}

We train our T2V diffusion model using the \textit{FIGR-8-SVG} dataset \cite{clouatre2019figr}, which consists of black-and-white vector icons. By eliminating stylistic variations, this dataset enables the model to focus on learning the structures and semantics of SVGs in the first stage.
To preprocess the data, we follow the same steps as IconShop \cite{wu2023iconshop} to obtain valid SVG data and their corresponding text descriptions. Examples from the dataset are shown in \reffig{vector_diff_samp}(a).
For the raw SVG data, we convert all other primitive shapes (\eg, lines, rectangles and ellipses) into cubic Bézier curves, which are encoded as path embeddings as described in \refsec{svg_representation}. Since the number of paths varies across SVGs, we pad the sequences of path tensors with zeros to a fixed length of 32 and filter out SVGs with more paths. This results in 210,000 samples for model training.

In our implementation, we set $d_P = 28$ and $m = 32$ for the SVG representation and normalize the SVG embeddings to the range of $[-1, 1]$. The model architecture consists of 28 transformer blocks, a hidden dimension of 800, and 12 attention heads. We configure the number of diffusion steps to $T = 1000$ and employ a cosine noise schedule \cite{ho2020denoising}. During training, we apply classifier-free guidance \cite{ho2022classifier} by randomly zeroing the text prompt $\mathbf{y}$ with a probability of 10\%. We use the Adam optimizer with an initial learning rate of $3 \times 10^{-5}$. The T2V diffusion network is trained for 3000 epochs with a batch size of 64, taking approximately 6 days on 8 A6000 GPUs.

After the first stage of training, our T2V diffusion model generates high-quality SVGs that align with text prompts while maintaining the structural integrity of the SVGs. In \reffig{vector_diff_samp}(b), we show several SVG samples generated by our model from random noise.

\section{Style Customization with Image Diffusion Priors}
\label{sec:style_customization_with_image_diffusion_priors}

In the second stage, we aim to enable style customization for the T2V diffusion model using only a few exemplar SVGs. A straightforward method of fine-tuning the T2V model with such a small dataset leads to overfitting. To address this issue, we distill style priors from different customized T2I models to generate a diverse set of customized images, which serve as augmented data for training the T2V model via an image-level loss.

\subsection{Style Distillation from Image Diffusion}
\label{sec:style_distillation_from_image_diffusion}

T2I diffusion models serve as a powerful prior for generating diverse images in customized styles. To enable the model to produce images in desired styles, we fine-tune a base T2I diffusion model "SD-v1-5" checkpoint\footnote{https://huggingface.co/runwayml/stable-diffusion-v1-5}, using a small set of style reference images. By applying the DreamBooth-LoRA method \cite{ruiz2022dreambooth, hu2021lora}, we create distinct LoRAs for each style. After fine-tuning, we concatenate a unique token $[V*]$ with the text prompt (\eg, "in $[V*]$ style") to generate customized images in the corresponding style using the specific LoRA.

Inspired by distillation techniques in T2I diffusion models \cite{luhman2021knowledge, yin2024one}, which generate noise-image pairs by running the sampling steps of the teacher diffusion model to train the student model, we apply a similar approach to generate customized images as guidance for training.
Given a text prompt $\mathbf{y}$, we randomly sample Gaussian noise $\boldsymbol{\epsilon}$ and input both into the T2V model. The T2V model performs DDPM denoising process to generate an SVG representation $\mathbf{s}_0^g$, which is then passed through a pre-trained path decoder \cite{zhang2024text} and a differentiable rasterizer $\mathcal{R}$ \cite{li2020differentiable} to produce an image $I_0^g=\mathcal{R}(Dec(\mathbf{s}_0^g))$. At this stage, $I_0^g$ is a black-and-white style image.
To ensure that the guidance image generated by the T2I model aligns with the structure of $I_0^g$ without significant deviations, we integrate ControlNet \cite{zhang2023adding} into the customized T2I model. This ensures overall structural alignment between the customized image $I_0^c$ and $I_0^g$. 
By using the Canny edge map of $I_0^g$ as a control image, we preserve the structural integrity of the original SVG while incorporating the desired style.
Using this approach, we can generate the corresponding $(\mathbf{s}_0^g, I_0^c)$ pair from random noise given a text prompt $\mathbf{y}$, and fine-tune the T2V model with image-level loss.

\subsection{Style Fine-tuning}
\label{sec:style_fine_tuning}

Given a generated SVG representation $\mathbf{s}_0^g$ and the corresponding customized image $I_0^c$, we fine-tune the T2V model towards new custom styles using image-level loss and diffusion loss.
In the forward diffusion process, Gaussian noise is added to $\mathbf{s}_0^g$, resulting in a noisy representation $\mathbf{s}_t^g$. We apply a reparameterization technique \cite{song2020score} to predict the denoised SVG tensor $\hat{\mathbf{s}}_0^g$ at each time step, by rewriting the closed-form sampling distribution for the forward diffusion process as: 
\begin{equation}
  \hat{\mathbf{s}}_0^g = (\mathbf{s}_t^g - \sqrt{1 - \bar{\alpha}_t} \cdot \boldsymbol{\epsilon}_\theta(\mathbf{s}_t^g, t)) / \sqrt{\bar{\alpha}_t}.
\end{equation}
By predicting $\hat{\mathbf{s}}_0^g$, we obtain the rendered image $\hat{I}_0^g = \mathcal{R}(\text{Dec}(\hat{\mathbf{s}}_0^g))$.
The image loss is computed as the MSE between the rendered image $\hat{I}_0^g$ and the customized image $I_0^c$:
\begin{equation}
  \mathcal{L}_{\text{img}} = \omega_t \| \hat{I}_0^g - I_0^c \|^2,
\end{equation}
where $\omega_t$ is a time-dependent weighting function designed to stabilize training by deactivating the loss term for noisier time steps. We set $\omega_t = (1 - \bar{\alpha}_t) $ empirically, following \cite{crowson2024scalable}.
The image loss guides the T2V model to predict SVGs that match the customized image, reflecting the desired style.
Additionally, we incorporate the diffusion loss $\mathcal{L}_{\text{DM}}$ defined on $\hat{\mathbf{s}}_0^g$, as described in Equation \ref{eq:diffusion_loss}, to help the model learn the new data distribution of the predicted SVGs $\hat{\mathbf{s}}_0^g$.
Specifically, Gaussian noise is added to $\hat{\mathbf{s}}_0^g$, and the model predicts this noise, with the diffusion loss calculated using Equation \ref{eq:diffusion_loss}.
During training, the T2V model is updated based on the combined loss function $\mathcal{L} = \mathcal{L}_{\text{img}} + \mathcal{L}_{\text{DM}}$.

To enable our T2V diffusion model to generate SVGs in diverse custom styles, we select 200 distinct style reference sets from SVGRepo\footnote{\url{https://www.svgrepo.com}}, iconfont\footnote{\url{https://www.iconfont.cn}} and Freepik\footnote{\url{https://www.freepik.com}}, each set containing a small collection of exemplar SVGs (ranging from 1 to 30 SVGs per set). We train the T2V model simultaneously across all 200 styles, with each style distinguished by a unique token "$[V*]$". The training process lasts for 80K iterations with a batch size of 20, where each iteration generates 20 pairs of $(\mathbf{s}_0^g, I_0^c)$ from randomly sampled text prompts and Gaussian noise across different styles. We employ a learning rate of $4 \times 10^{-6}$, taking approximately 6 days using 8 A6000 GPUs.
After style fine-tuning, our T2V model can generate SVGs in the learned custom styles based on text prompts in a feed-forward manner.

Our method also supports fine-tuning on a single style or incrementally adding a new style with only a few exemplars via either full-model or LoRA fine-tuning. 
Similar to DreamBooth \cite{ruiz2022dreambooth}, the former approach fine-tunes the full T2V model with a new style represented by a token "$[V*]$" that is distinct from all existing style tokens.
For a parameter-efficient alternative, we fine-tune external LoRA weights for the attention layers of DiT blocks \cite{hu2021lora}, adapting the model to a new style without introducing an additional style token.

\section{Experiments}
\label{sec:experiments}

\paragraph{Experiment Setup}

To evaluate our method, we randomly select 5 text prompts from the \textit{FIGR-8-SVG} dataset \cite{clouatre2019figr} for each of the 200 styles, resulting in a total of 1000 vector graphics. For each style, we append the special style token "in $[V*]$ style" to the respective text prompt.
During testing, we use the DDPM sampler with 1000 steps and classifier-free guidance with a scale of 3 to achieve better results. Generating an SVG takes around 25 seconds on an NVIDIA-A6000.

\paragraph{Evaluation Metrics}
\label{sec:evaluation_metrics}

We evaluate the quality of our results from vector-level, image-level, text-level perspectives.
For \textbf{vector-level} evaluation, we use a path VAE \cite{zhang2024text} trained on the \textit{FIGR-8-SVG} dataset to encode SVG paths into latent vectors. We calculate the FID between these latents and the ground truth paths from \textit{FIGR-8-SVG}, to evaluate how well the paths align with well-designed vector graphics.
For \textbf{image-level} evaluation, we evaluate the style alignment and the visual aesthetics of SVGs. We measure style alignment by calculating the average cosine similarity between ClIP image features of style references and rendered SVG images \cite{sohn2023styledrop}. We use the Aesthetic score \cite{schuhmann2021improved} to evaluate the overall image quality.
For \textbf{text-level} evaluation, we calculate the CLIP cosine similarity \cite{radford2021learning} between the text prompt and the rendered SVGs to measure semantic alignment.

\paragraph{Baselines}
\label{sec:baseslines}

We compare our proposed pipeline with two types of T2V generation schemes: optimization-based methods and feed-forward methods.

\textbf{Optimization-based methods} rely on pre-trained T2I models, so we first perform style-tuning on T2I diffusion models using the method described in \refsec{style_distillation_from_image_diffusion}.
For vectorization with T2I methods, we generate customized images and optimize the SVGs to fit the images using two distinct vectorization techniques: a traditional method, Potrace \cite{selinger2003potrace}, and a deep learning-based method LIVE \cite{ma2022towards}.
For text-guided SVG optimization, we compare three approaches: VectorFusion \cite{jain2022vectorfusion} using SDS loss, SVGDreamer \cite{xing2023svgdreamer} and T2I-NPR \cite{zhang2024text}, which use VSD loss. We use 64 paths for SVG optimization. To enhance alignment with the exemplar style, we begin by using the vectorized outputs from customized images as the initial SVGs.

\textbf{Feed-forward methods} include language-based and diffusion-based approaches.
For the former, we use GPT-4o \cite{achiam2023gpt} to generate customized SVGs via providing curated in-context examples in the prompts. Specifically, we supply raster images and corresponding SVG scripts of the exemplar style and let GPT-4o act as an SVG code generator, to generate SVGs that match the style of the exemplars with the given text prompts.
For the latter, since no diffusion-based T2V models are publicly available yet, we reproduce VecFusion \cite{thamizharasan2024vecfusion} as a base T2V model. To achieve style customization, we compare two approaches: (1) an vector-based fine-tuning method, in which we fine-tune VecFusion with a small set of exemplar SVGs \cite{ruiz2022dreambooth}; (2) a neural style transfer (NST) method for SVG \cite{efimova2023neural}, where we first generate an SVG using the base model, then select an exemplar SVG as the style reference and apply style transfer to the model's output.

\subsection{Comparisons}

\begin{figure}[tbp]
  \centering
  \includegraphics[width=1.0\columnwidth]{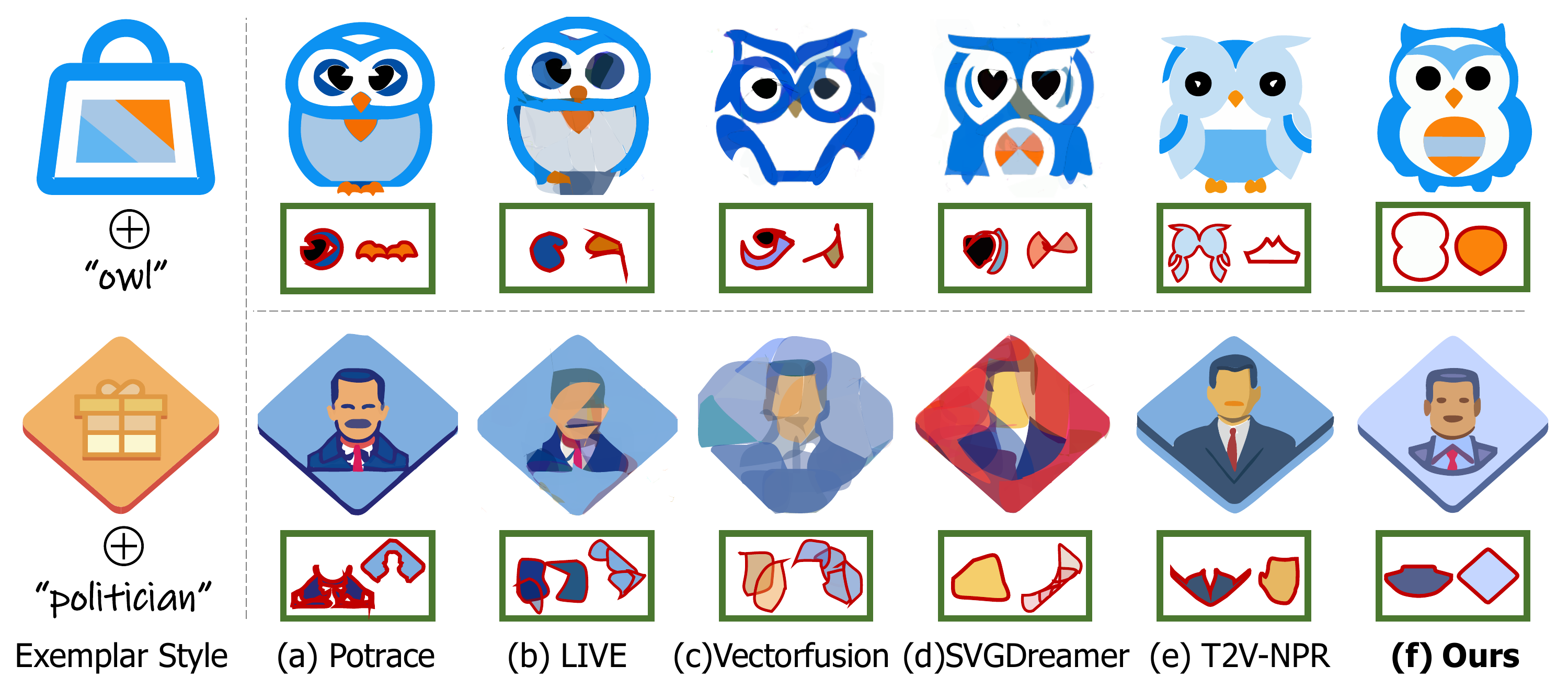}
  \caption{ \label{fig:result_optm1} Qualitative comparison with optimization-based T2V methods. Exemplar SVGs are from \copyright{SVGRepo}. }
\end{figure}

\begin{figure}[tbp]
  \centering
  \includegraphics[width=1.0\columnwidth]{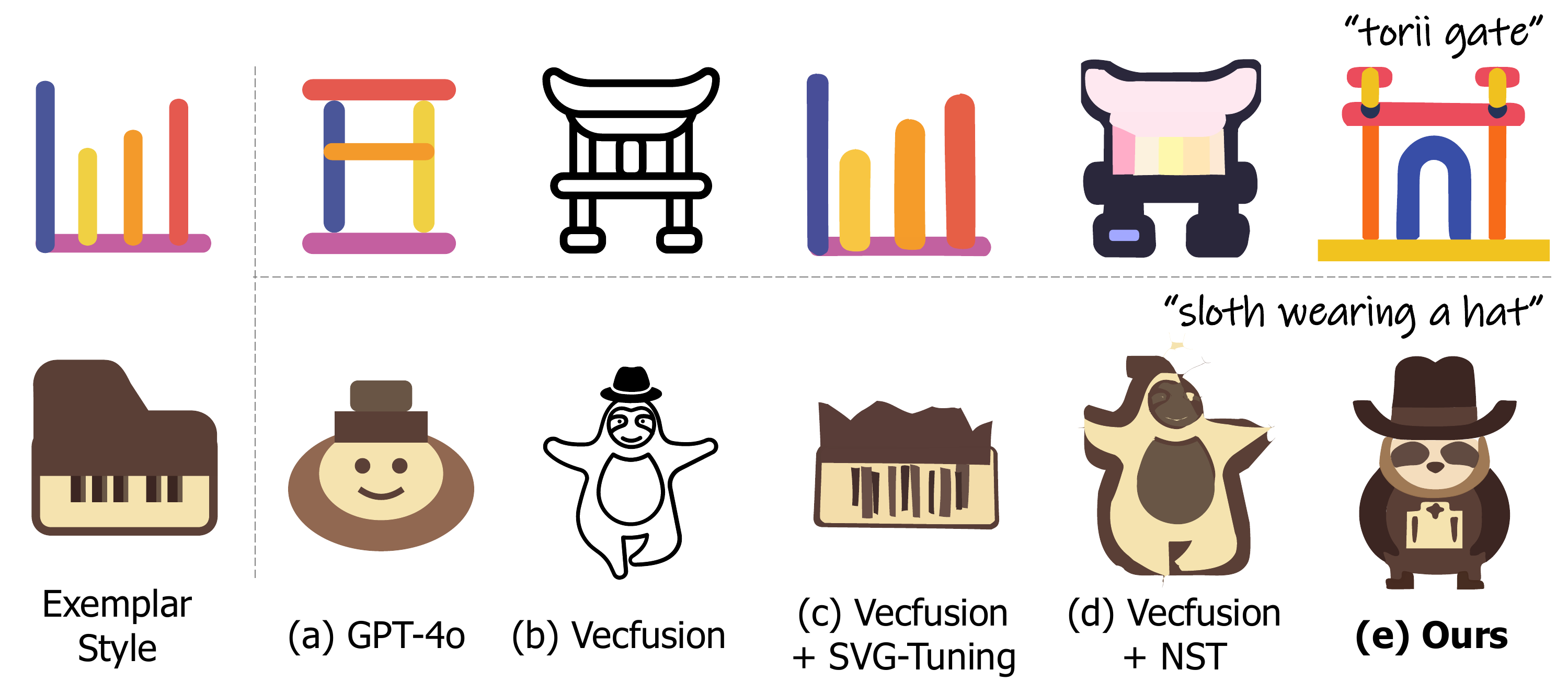}
  \caption{ \label{fig:result_ff1} Qualitative comparison to feed-forward T2V methods. Exemplar SVGs are from \copyright{SVGRepo}. }
\end{figure}

\begin{figure*}[tbp]
  \centering
  \includegraphics[width=0.91\linewidth]{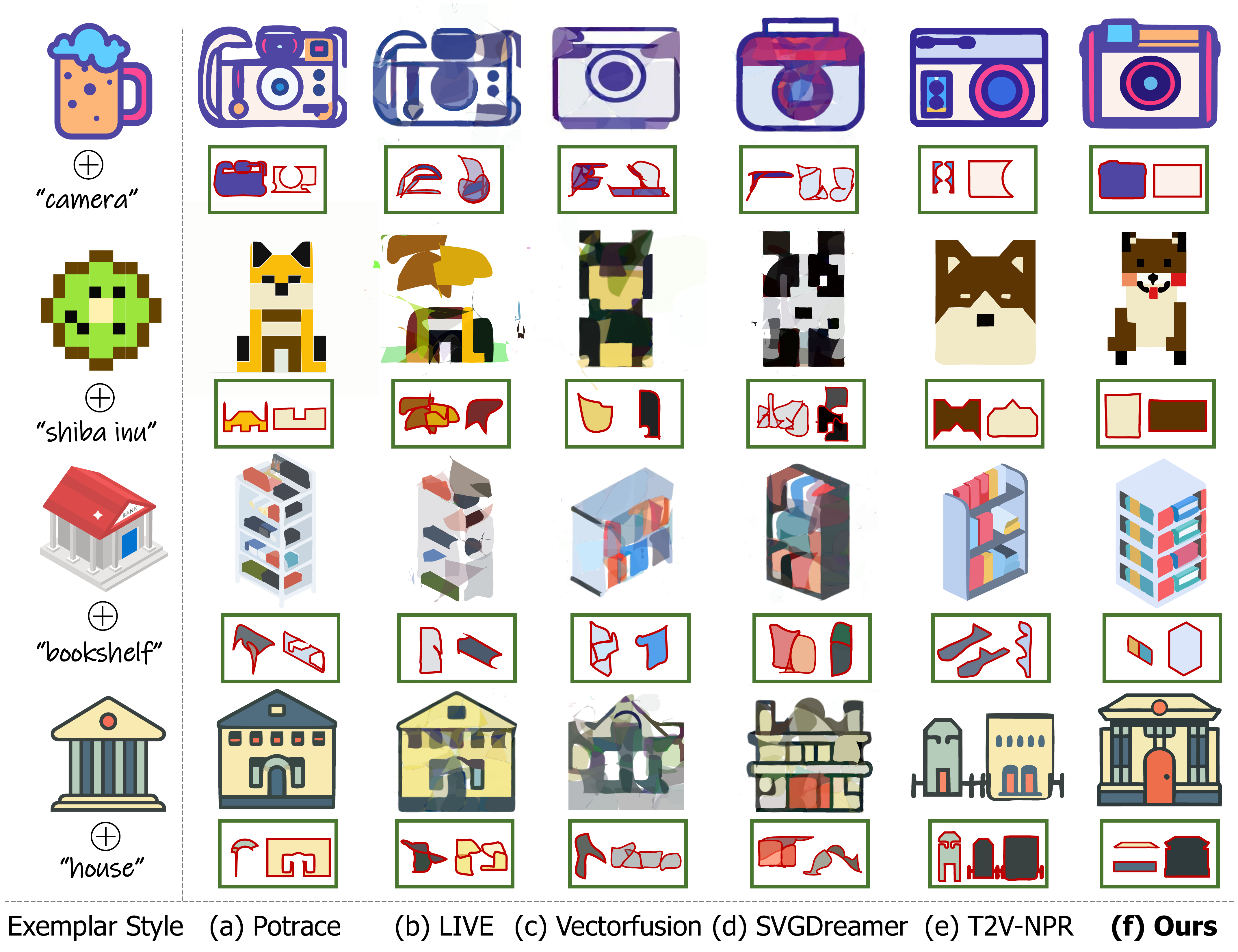}
  \caption{ \label{fig:result_optm2} More qualitative comparison with optimization-based T2V methods. Exemplar SVGs: the $1^{st}$, $2^{nd}$ and $4^{th}$ rows are from \copyright{SVGRepo}; the $3^{rd}$ row is from \copyright{Freepik}. }
\end{figure*}

\begin{figure*}[tbp]
  \centering
  \includegraphics[width=0.7\linewidth]{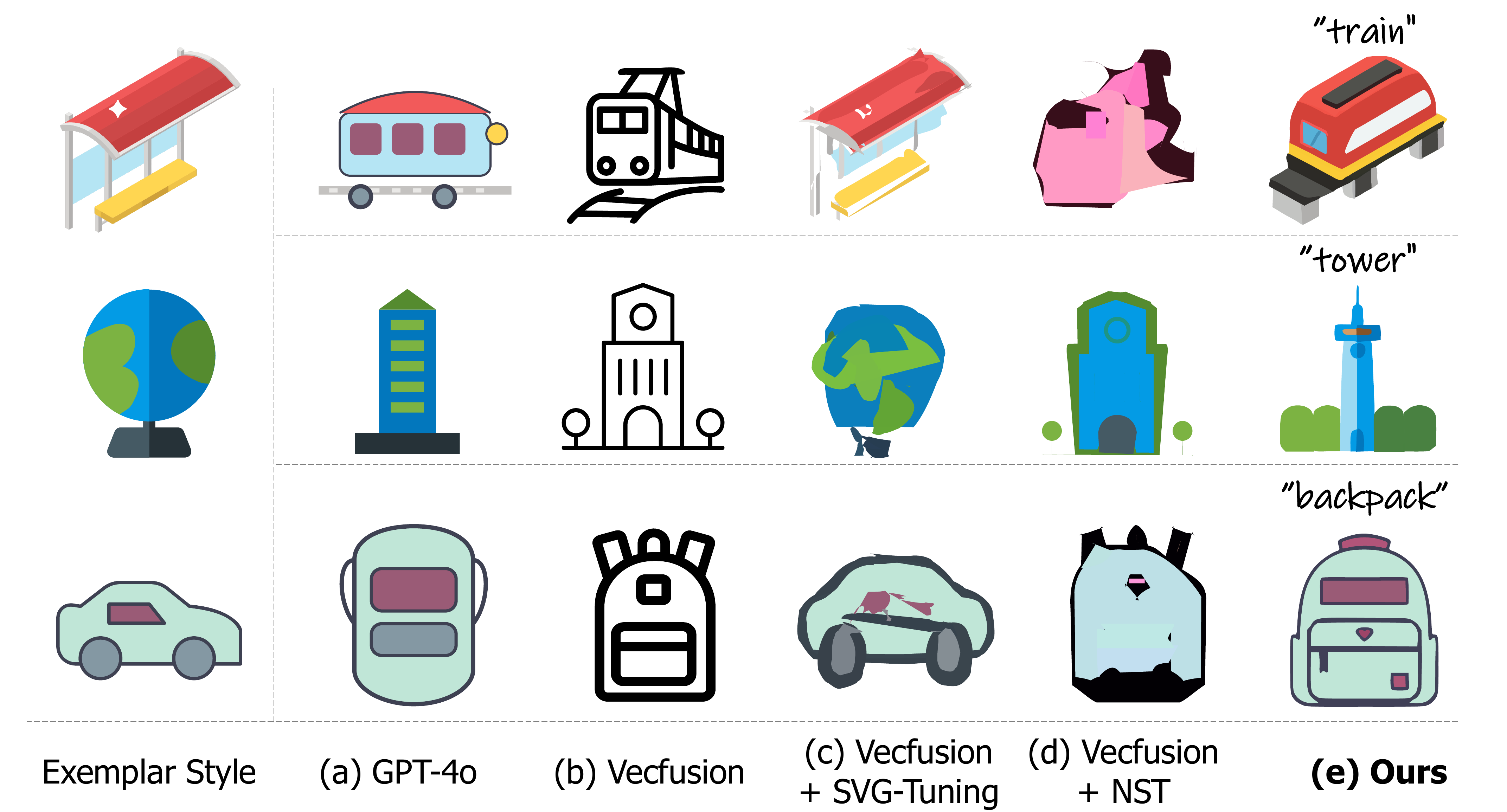}
  \caption{ \label{fig:result_ff2} More qualitative comparison to feed-forward T2V methods. Exemplar SVGs: the $1^{st}$ row is from \copyright{Freepik}; the $2^{nd}$ and $3^{rd}$ rows are from \copyright{SVGRepo}. }
\end{figure*}

We evaluate the performance of our method by comparing it with baselines qualitatively and quantitatively. The quantitative results are provided in \reftab{table_quality_eval} and the qualitative results are presented in \reffig{result_optm1}, \reffig{result_ff1}, \reffig{result_optm2} and \reffig{result_ff2}. As shown in \reftab{table_quality_eval}, our method outperforms the others from a comprehensive perspective.

\paragraph{Comparisons with Optimization-based Methods}

Vectorization with T2I methods reconstruct customized images through color-based image segmentation and curve fitting.
However, as shown in \reffig{result_optm1} and \reffig{result_optm2}(c), while Potrace \cite{selinger2003potrace} produces visually appealing outputs by faithfully reconstructing the customized images, it struggles with overly complex vector elements and lacks layer organization, as indicated by the higher Path FID in \reftab{table_quality_eval}. This results in disorganized structures, reduced semantic clarity, and increased complexity in the SVGs. Such issues are common in image vectorization methods and contradict professional design principles, which prioritize simplicity and clarity in vector graphics.
LIVE \cite{ma2022towards} faces similar challenges, producing SVGs containing numerous irregular and broken paths. Zoomed-in illustrations in \reffig{result_optm2} highlight the issues of overcomplicated and fragmented paths (shown within green boxes).

Text-guided SVG optimization methods leverage score distillation in T2I diffusion models to optimize a set of shapes.
VectorFusion \cite{jain2022vectorfusion} and SVGDreamer \cite{xing2023svgdreamer} directly optimize the control points of paths to generate text-conforming SVGs. However, due to the high degrees of freedom, the paths may undergo complex transformations that result in jagged and cluttered shapes, leading to visually unappealing outcomes.
T2V-NPR \cite{zhang2024text} tackles the issue of irregular paths by learning a latent representation of paths and reduces fragmentation by merging shapes with similar colors. However, while it produces smooth paths, it cannot guarantee the semantic integrity of the shapes, as the merging operation overlooks their semantic meaning. This can lead to semantically ambiguous paths, such as the merging of an owl's eye with its wing, as shown in the first row of \reffig{result_optm1}.

Overall, optimization-based methods that rely only on image supervision overlook the inherent design principles and layer structure of SVGs. Consequently, the generated SVGs often contain redundant shapes and disorganized layers, making them difficult to edit.
Moreover, these methods are time-consuming due to their iterative optimization, typically requiring tens of minutes per SVG, which makes them impractical for real design scenarios.
In contrast, our T2V diffusion model learns vector properties, such as valid path semantics and layer structure, by training on a well-designed SVG dataset. Our novel two-stage training strategy enables feed-forward generation of well-structured SVGs in a few seconds, while achieving visually appealing results in diverse custom styles.

\begin{table}[tbp]
  \caption{ Quantitative comparison with existing methods. }
  \resizebox{\linewidth}{!}{
    \begin{tabular}{|cc|c|c|c|c|}
      \hline
      \multicolumn{2}{|c|}{Methods}                                                                     & \begin{tabular}[c]{@{}c@{}}Path \\ FID\end{tabular} $\big\downarrow$ & \begin{tabular}[c]{@{}c@{}}Style \\ Alignment\end{tabular} $\big\uparrow$ & \begin{tabular}[c]{@{}c@{}}Visual\\ Aesthetic\end{tabular} $\big\uparrow$ & \begin{tabular}[c]{@{}c@{}}Text \\ Alignment\end{tabular} $\big\uparrow$ \\ \hline
      \multicolumn{1}{|c|}{\multirow{5}{*}{Optimization}} & Potrace                                     & 44.29                 & 0.665                      & 5.522                & 0.294                     \\ \cline{2-6} 
      \multicolumn{1}{|c|}{}                              & LIVE                                        & 52.43                 & 0.578                      & 4.686                & 0.258                     \\ \cline{2-6} 
      \multicolumn{1}{|c|}{}                              & Vectorfusion                                & 53.76                 & 0.557                      & 4.892                & 0.276                     \\ \cline{2-6} 
      \multicolumn{1}{|c|}{}                              & SVGDreamer                                  & 48.51                 & 0.564                      & 5.013                & 0.281                     \\ \cline{2-6} 
      \multicolumn{1}{|c|}{}                              & T2V-NPR                                     & 40.25                 & 0.608                      & 5.237                & 0.290                     \\ \hline
      \multicolumn{1}{|c|}{\multirow{4}{*}{Feed-forward}} & GPT-4o                                      & 38.14                 & 0.549                      & 5.041                & 0.251                     \\ \cline{2-6} 
      \multicolumn{1}{|c|}{}                              & \multicolumn{1}{l|}{VecF + SVG-FT} & 45.05                 & \textbf{0.726}             & 4.980                & 0.223                     \\ \cline{2-6} 
      \multicolumn{1}{|c|}{}                              & VecF + NST                             & 58.12                 & 0.573                      & 4.574                & 0.245                     \\ \cline{2-6} 
      \multicolumn{1}{|c|}{}                              & \textbf{Ours}                               & \textbf{37.51}        & 0.661                      & \textbf{5.527}       & \textbf{0.297}            \\ \hline
      \end{tabular}
  }
  \label{tab:table_quality_eval}
\end{table}

\paragraph{Comparisons with Feed-forward Methods}

Regarding language-based methods, though provided with in-context SVG examples, GPT-4o \cite{achiam2023gpt} can only generate simple combinations of basic primitive shapes (\eg, circles and rectangles) to align with text prompts, as illustrated in \reffig{result_ff1}(a). It fails to produce the complex geometric details required for professional SVGs, resulting in outputs inadequate for graphic design.

Although the original VecFusion model \cite{thamizharasan2024vecfusion}, trained on the \textit{FIGR-8-SVG} dataset, generates high-quality results within its trained domains, it cannot be extended to style customization using existing methods. When applying an vector-based fine-tuning approach, where VecFusion is fine-tuned with a small set of exemplar SVGs \cite{ruiz2022dreambooth}, its limited generalization ability prevents it from generating semantically correct SVGs in new custom styles. Instead, the model overfits to the exemplar SVGs, simply reconstructing them rather than adapting to diverse prompts. As a result, the generated outputs exhibit high Style Alignment but poor Text Alignment in \reftab{table_quality_eval}.

NST \cite{efimova2023neural} applies style transfer to the SVGs generated by VecFusion using a style loss in image space. Although this method directly inherits the original layer-wise properties, the optimized SVGs often have messy visual appearances. Furthermore, it struggles to capture fine-grained style features, leading to poor style consistency.

In contrast, our method excels at adapting to the style, effectively capturing details from user-provided style, such as color schemes and design patterns. It achieves high visual quality while preserving the structure of the output SVGs.

\subsection{User Study}
\label{sec:user_study}

\begin{figure}[tbp]
  \centering
  \includegraphics[width=\columnwidth]{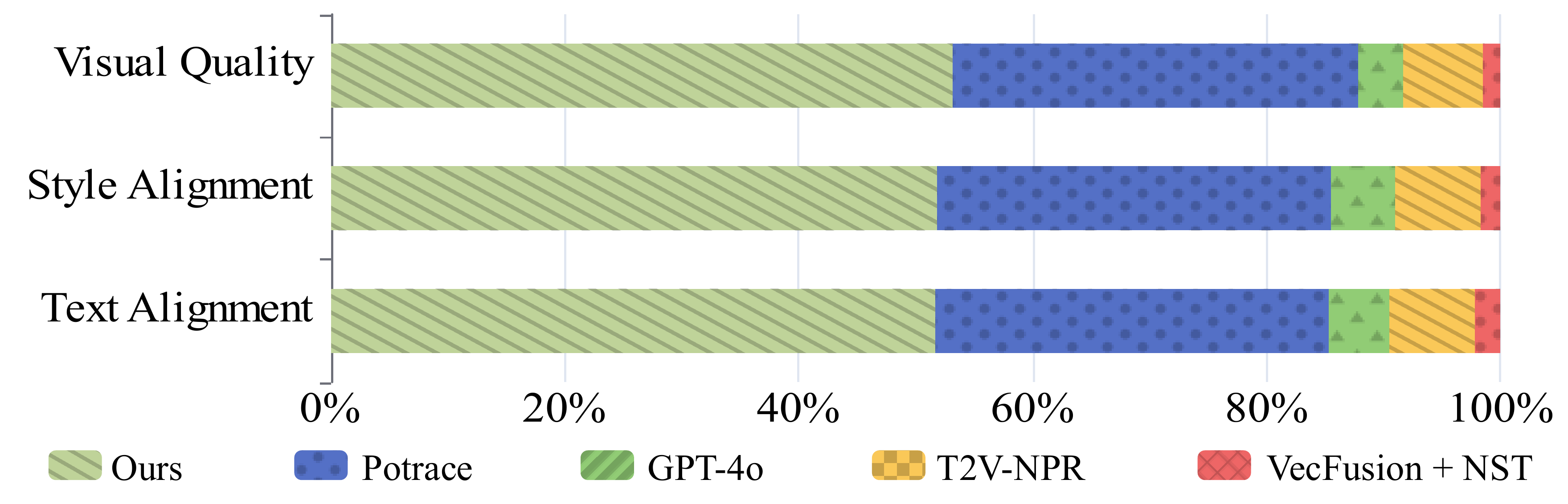}
  \caption{ \label{fig:user_study_bar} User Study. We show the human preferences in \%. }
\end{figure}

We conducted a perceptual study to evaluate our style customization of T2V generation from three perspectives: overall SVG quality, style alignment and semantic alignment.
We randomly selected 20 text prompts from the dataset and generated SVGs using both the baseline methods and our approach.
Each question presented the results of different methods in a random order, and 30 participants were given unlimited time to select the best result among five options for each evaluation metric.
\reffig{user_study_bar} demonstrates the superior performance of our method, as it achieves the highest preference in all evaluation metrics. Specifically, our method obtains 53.2\% of votes for overall SVG quality, 51.8\% for style alignment, and 51.7\% for semantic alignment. The results show the effectiveness of our method in generating high-quality SVGs in custom styles from text prompts that align more closely with human perception.

\subsection{Ablation Study}
\label{sec:ablation_study}

\paragraph{Ablation on SVG Representation}

\begin{figure}[tbp]
  \centering
  \includegraphics[width=\columnwidth]{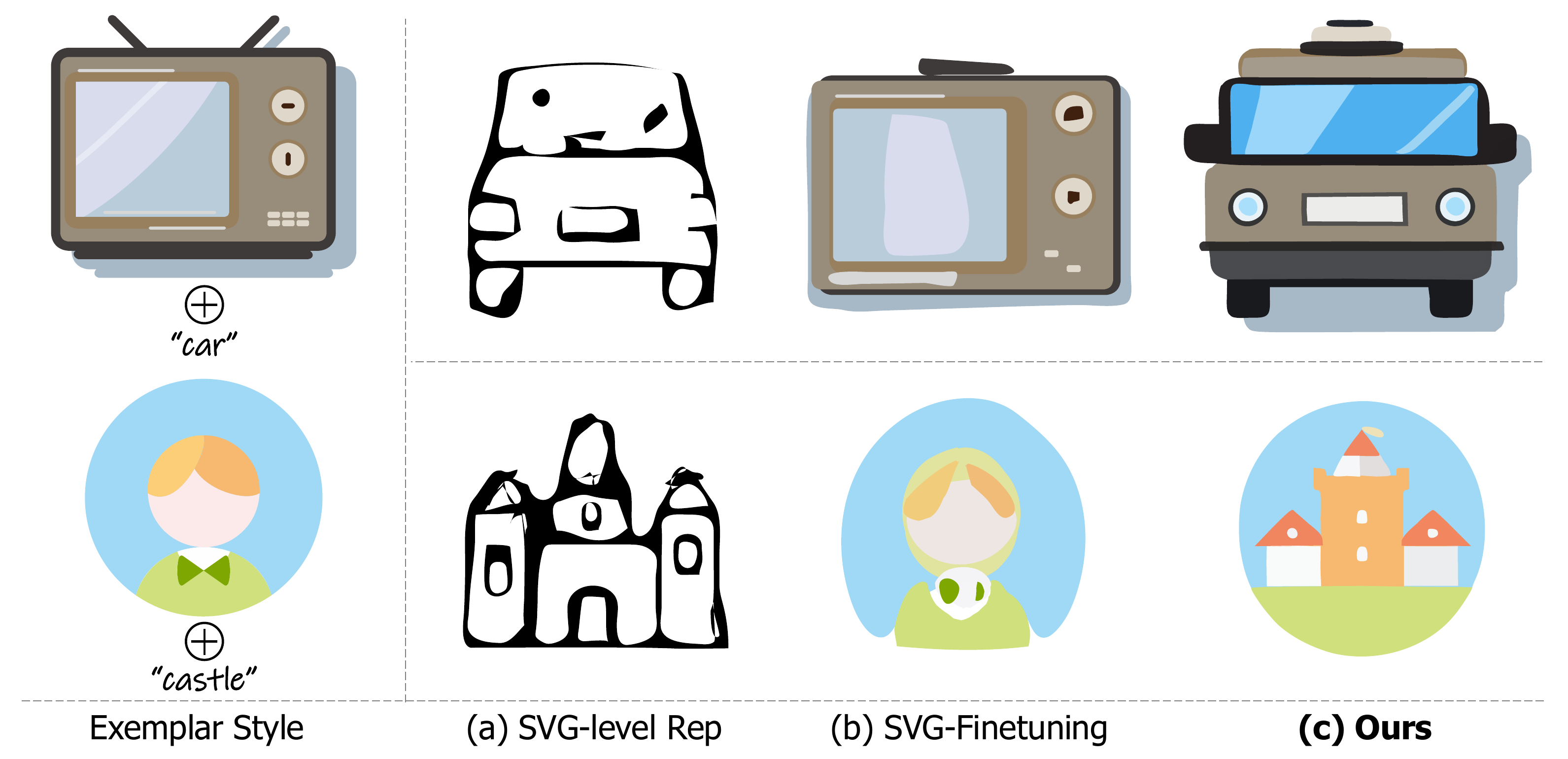}
  \caption{\label{fig:ablation} Qualitative results on ablation study. Exemplar SVGs are from \copyright{SVGRepo}. }
\end{figure}

Instead of using our path-level representation for training the T2V diffusion model, another baseline is to use a global SVG-level representation.
Following the Deep-SVG \cite{carlier2020deepsvg} architecture, we train a transformer-based VAE on the \textit{FIGR-8-SVG} dataset, where all paths with their properties (including the path order, control points and color) are encoded into a single latent vector. We then replace our path-level representation with this SVG-level representation for T2V diffusion model training and subsequent style distillation.
However, the global SVG-level representation is constrained by the geometry and color limitations of the dataset, which restricts its ability to generate SVGs in only a fixed style. As a result, it fails to adapt to new custom styles, as shown in \reffig{ablation}(a). In contrast, our path-level representation maintains both compactness and expressivity, allowing for flexible and diverse SVG customizations.

\paragraph{Ablation on Style Customization with Image Diffusion Priors}

We compare our image-based style customization method with a vector-based fine-tuning approach. Specifically, in the second stage, we directly fine-tune our T2V model using a small set of exemplar SVGs, following the fine-tuning techniques of T2I diffusion models \cite{ruiz2022dreambooth}. However, as shown in \reffig{ablation}(b), this method leads to overfitting on the exemplar SVGs, causing the model to simply reconstruct them rather than aligning with the text semantics, as reflected by a high style alignment score and a low text alignment score in \reftab{ablation_study}. In contrast, our style distillation method from image diffusion takes advantage of the strong visual priors in T2I diffusion models to generate customized images as augmented data, enabling diverse style customizations of SVGs.

\begin{table}[tbp]
  \caption{ Ablation study on SVG representation and style customization with image diffusion priors. }
  \resizebox{\linewidth}{!}{
    \begin{tabular}{|c|c|c|c|c|}
      \hline
      {Methods}                                                                     & \begin{tabular}[c]{@{}c@{}}Path \\ FID\end{tabular} $\big\downarrow$ & \begin{tabular}[c]{@{}c@{}}Style \\ Alignment\end{tabular} $\big\uparrow$ & \begin{tabular}[c]{@{}c@{}}Visual\\ Aesthetic\end{tabular} $\big\uparrow$ & \begin{tabular}[c]{@{}c@{}}Text \\ Alignment\end{tabular} $\big\uparrow$ \\ \hline
      SVG-level Rep & 45.16                 & 0.406                      & 3.285                & 0.288                     \\ \hline
      SVG-FT           & 57.32                 & \textbf{0.722}                      & 4.926                & 0.221                     \\ \hline
      \textbf{Ours}            & \textbf{37.51}        & 0.661                      & \textbf{5.527}       & \textbf{0.297}            \\ \hline
      \end{tabular}
  }
  \label{tab:ablation_study}
\end{table}

\section{Conclusion}
\label{sec:conclusion}

\begin{figure}[tbp]
  \centering
  \includegraphics[width=1.0\columnwidth]{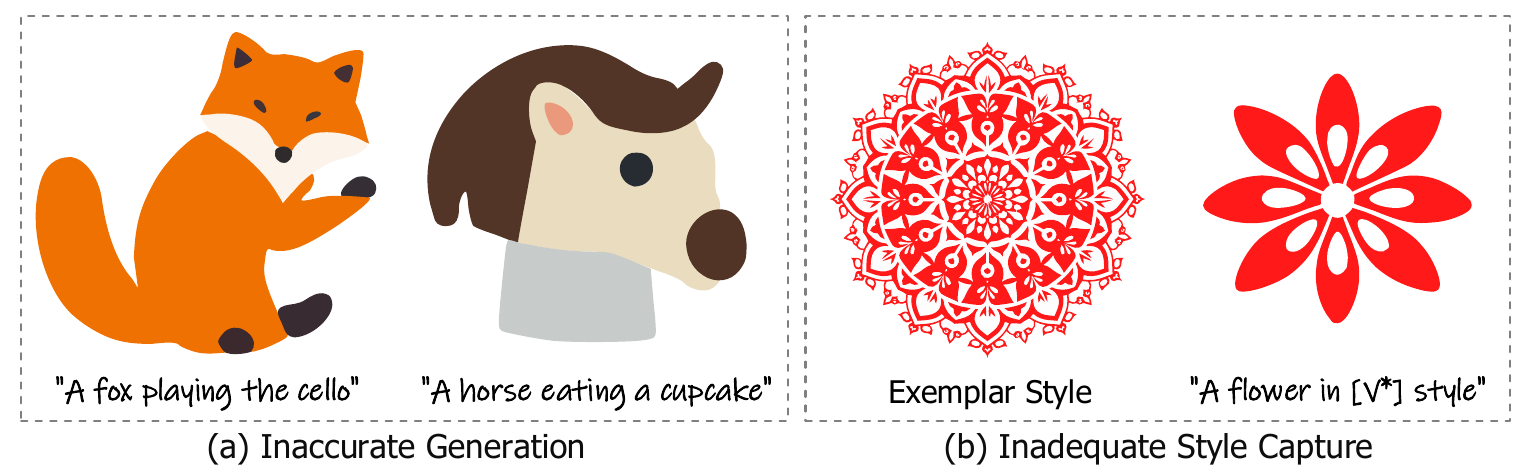}
  \caption{ \label{fig:failure_case} Failure cases. The exemplar SVG is from \copyright{iconfont}. }
\end{figure}

In this paper, we present a novel two-stage pipeline for style customization of SVGs. Our approach disentangles content and style semantics in the T2V diffusion model, ensuring structural regularity and expressive diversity in the generated SVGs. By employing a path-level T2V diffusion model and distilling styles from T2I diffusion priors, our method produces high-quality SVGs in custom styles from text prompts in a feed-forward manner.
While our method excels at SVG style customization, it has limitations. First, our T2V model is trained on the \textit{FIGR-8-SVG} dataset, which contains only simple class labels, limiting the model's semantic understanding of SVG content. For example, as shown in \reffig{failure_case}(a), semantic elements like "cello" and "cupcake" are inaccurate when the text descriptions exceed the training domain's capacity. This could be mitigated with a larger and higher-quality SVG dataset with detailed annotations. Second, it may lose fine-grained stylistic details for overly complex style references, as depicted in \reffig{failure_case}(b).
Our model can be used to synthesize SVG data, and with advanced diffusion model techniques, it enables flexible control and editing, which we plan to explore in future work.

\begin{acks}
The work described in this paper was substantially supported by a GRF grant from the Research Grants Council (RGC) of the Hong Kong Special Administrative Region, China [Project No. CityU 11216122].
\end{acks}

\bibliographystyle{ACM-Reference-Format}
\bibliography{sample-base}

\begin{figure*}[tbp]
  \centering
  \includegraphics[width=1.0\linewidth]{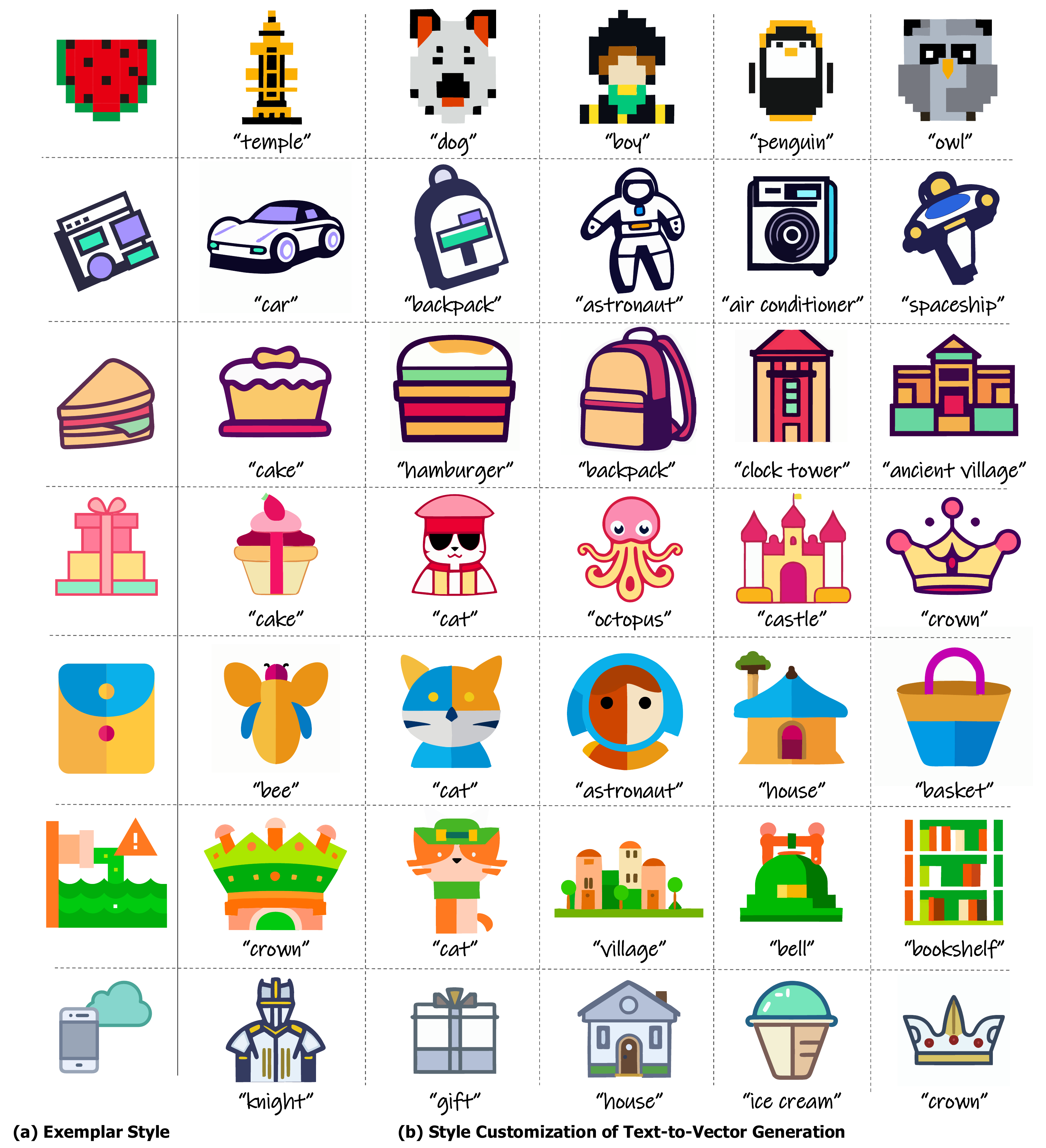}
  \caption{\label{fig:more_results} More results of our style customization of T2V generation. Exemplar SVGs: the $1^{st}$, $2^{nd}$, $3^{rd}$, $4^{th}$, $5^{th}$ and $7^{th}$ rows are from \copyright{SVGRepo}; the $6^{th}$ row is from \copyright{Freepik}. }
\end{figure*}

\end{document}